# An antisite defect mechanism for room temperature ferroelectricity in orthoferrites

Shuai Ning[1*], Abinash Kumar[1], Konstantin Klyukin[1], Jong Heon Kim[2], Tingyu Su[1], Hyun-Suk Kim[2], James M. LeBeau[1], Bilge Yildiz[1,3] and Caroline A. Ross[1*]

[1]Department of Materials Science and Engineering, Massachusetts Institute of Technology, Cambridge, Massachusetts 02139, USA.

[2]Department of Materials Science and Engineering, Chungnam National University, Daejeon 305-764, Korea

[3]Department of Nuclear Science and Engineering, Massachusetts Institute of Technology, Cambridge, Massachusetts 02139, USA.

*E-mail: sning@mit.edu, caross@mit.edu

**Abstract**

Single-phase multiferroic materials that allow the coexistence of ferroelectric and magnetic ordering above room temperature are highly desirable, motivating an ongoing search for mechanisms for unconventional ferroelectricity in magnetic oxides. Here, we report an antisite defect mechanism for room temperature ferroelectricity in epitaxial thin films of yttrium orthoferrite, $YFeO_3$, a perovskite-structured canted antiferromagnet. A combination of piezoresponse force microscopy, atomically resolved elemental mapping with aberration corrected scanning transmission electron microscopy and density functional theory calculations reveals that the presence of $Y_{Fe}$ antisite defects facilitates a non-centrosymmetric distortion promoting ferroelectricity. This mechanism is predicted to work analogously for other rare earth orthoferrites, with a dependence of the polarization on the radius of the rare earth cation. Furthermore, a vertically aligned nanocomposite consisting of pillars of a magnetoelastic oxide $CoFe_2O_4$ embedded epitaxially in the $YFeO_3$ matrix exhibits both robust ferroelectricity and ferrimagnetism at room temperature, as well as a noticeable strain-mediated magnetoelectric coupling effect. Our work uncovers the distinctive role of antisite defects in providing a novel mechanism for ferroelectricity in a range of magnetic orthoferrites and further augments the functionality of this family of complex oxides for multiferroic applications.

**Introduction**

Multiferroic materials combine ferroelectric and magnetic properties and hold compelling interest for magnetoelectric and spintronic applications[1]. It has traditionally been a challenge for ferroelectricity and magnetism to coexist in a single-phase material, particularly the perovskite-structured ($ABO_3$) transition metal oxides[2]. While magnetoelectric coupling can be artificially realized by integrating ferroelectric and magnetic phases into composites or heterostructures[3], single-phase multiferroic materials and mechanisms offer substantial fundamental and technological interest and are highly desirable for practical applications.

One of the best-known single-phase multiferroics is BiFeO$_3$ (BFO)[4], which possesses large spontaneous polarization (~100 μC/cm$^2$) driven by the stereochemical activity of lone-pair electrons of *A*-site Bi$^{3+}$ ions and weak magnetization due to the canted antiferromagnetic ordering of *B*-site Fe$^{3+}$ ions[5,6]. Beyond this, novel mechanisms for multiferroicity have been established by incorporating unconventional ferroelectricity in magnetic oxides[7], such as the geometrically-induced ferroelectricity due to polyhedral tilt and rotation in hexagonal manganites[8,9] and ferrites[10,11]; electronically-driven ferroelectricity due to charge ordering[12,13]; and magnetically-induced ferroelectricity due to types of non-centrosymmetric spin ordering[14]. Strain engineering is also predicted to promote ferroelectricity in certain magnetic oxides, *e.g.* EuTiO$_3$[15,16] and SrMnO$_3$[17,18], by spin-lattice coupling.

Among such approaches, magnetically-driven ferroelectricity is of great interest due to the inherent coupling expected between the ferroelectric and magnetic orderings. It was first reported in TbMnO$_3$ that inversion symmetry breaking is caused by a cycloidal spiral spin structure at cryogenic temperature[19]. Subsequently, ferroelectricity with a weak polarization (0.1 μC/cm$^2$) was also observed in a family of rare earth orthoferrites (*R*FeO$_3$) with much simpler collinear spin structures, *e.g.* GdFeO$_3$[20] and DyFeO$_3$[21]. The mechanism is attributed to the exchange-striction effect of *R* 4*f* and Fe 3*d* spins, and hence the ferroelectricity only occurs below the *R* 4*f* spin ordering temperature of a few Kelvin. In HoFeO$_3$[22] and SmFeO$_3$[23], ferroelectricity is reported to be present near or even above room temperature, but contradictory conclusions from theoretical[24] and experimental[25,26] work have obscured the mechanism, particularly in the case of room temperature ferroelectricity reported in YFeO$_3$ (YFO)[27,28,29]. YFO adopts a space group *Pbnm* with lattice parameters of $a_o$=5.282 Å, $b_o$=5.595 Å, $c_o$=7.605 Å (the subscript o denotes the orthorhombic notation). Such a centrosymmetric structure inhibits spontaneous polarization in principle. The unexpected ferroelectricity found in YFO was initially attributed to spin-orbit-coupling[27], *i.e.* the inverse Dzyaloshinskii-Moriya interaction (DMI), as proposed for SmFeO$_3$[23]. However, it was shown that inverse DMI is unable to break the inversion symmetry in SmFeO$_3$[24], precluding this mechanism. The ferroelectricity also cannot stem from spin exchange interactions because of the empty 4*f* orbital of Y$^{3+}$ ions. Moreover, the reported polarization of YFO varies dramatically in magnitude from bulk (<0.01 μC/cm$^2$)[27] to one thin film report (~10 μC/cm$^2$)[28], challenging the interpretations of the origin of its polarization.

In this work, we select YFO as a model system to demonstrate an antisite defect mechanism that yields a robust, sizeable and switchable ferroelectric polarization above room temperature while preserving its magnetization. Contrary to the fact that the ferroelectricity generally deteriorates upon the presence of cationic off-stoichiometry in incipient ferroelectric CaTiO$_3$[30], and analogous to the findings that cationic off-stoichiometry yields a ferroelectric distortion in antiferroelectric PbZrO$_3$[31] or paraelectric SrTiO$_3$[32,33], the Y-rich composition, specifically the presence of Y-Fe antisite (Y$_{Fe}$) defects, plays a crucial role in facilitating a non-centrosymmetric distortion which promotes a spontaneous polarization but preserves the magnetic order. The polarization persists over a range of film thicknesses and lattice strains, and is supported by both theoretical simulations and experimental characterization. Density functional theory (DFT) calculations demonstrate that such an antisite defect mechanism is also expected for other rare earth orthoferrites in which the polarization depends on the radius of the rare earth cation. Furthermore, we succeed in integrating YFO with magnetoelastic CoFe$_2$O$_4$ (CFO) to form a vertically aligned epitaxial nanocomposite, where a noticeable strain-mediated magnetoelectric coupling effect is observed. Overall, our results demonstrate a cationic antisite defect mechanism for ferroelectricity in the family of orthoferrites, offering a design strategy for novel single-phase multiferroics.

**Results**

*Unexpected ferroelectricity in magnetic YFO thin films*

YFO thin films (~30 nm thick, Supplementary Note 1) were grown on 001-oriented Nb-doped SrTiO$_3$ (NSTO) ($a$=3.905 Å) substrates by pulsed laser deposition (PLD, see Methods) using a stoichiometric YFeO$_3$ ceramic target. High-resolution X-ray diffraction (XRD, Fig. 1a) reveals that the YFO film exhibits an epitaxial perovskite structure, while the asymmetric (013) reciprocal space mapping (RSM, Fig. 1b) shows it is not fully strained to the substrate. The broadening of the RSM peak suggests that imperfections such as mosaicity exist. With the YFO lattice described using a pseudocubic unit cell (Fig. 1c), the in-plane ($a_p$, the subscript p denotes the pseudocubic notation) and out-of-plane ($c_p$) lattice parameters of the YFO film are 3.862 Å and 3.813 Å, respectively, both of which are larger than the bulk values ($a_{p,\,bulk}=\frac{1}{2}\sqrt{a_o^2+b_o^2}$=3.847 Å, $c_{p,\,bulk}=c_o/2$=3.803 Å), *i.e.* the unit cell volume of the thin film is larger by +1.04% compared to bulk. The chemical composition analyzed by high resolution X-ray photoelectron spectroscopy (XPS) reveals that the as-prepared YFO film possesses a Y-rich composition, which is surprisingly off-stoichiometric compared to the target (Fig. 1d) and consistent across multiple samples prepared during several PLD runs under the same deposition conditions, which yielded a Y/Fe ratio ($\alpha$) of 1.19 ± 0.04 (Supplementary Note 1). STEM EDS analysis also reveals a uniform composition with Y/Fe ratio of 1.21 ± 0.1, further confirming the Y-rich nature of the films (Supplementary Note 1). This cationic off-stoichiometry can arise from differences in ablation or scattering rate of Y and Fe during the PLD process[34].

Atomic-resolved scanning transmission electron microscopy (STEM) energy dispersive spectroscopy (EDS) elemental mapping (Fig. 1e) on the YFO/NSTO with $α=1.19$ reveals the presence of Y ions at Fe sites, *i.e.* $Y_{Fe}$ defects. Based on the composition and the lamella thickness (11 nm), 2-3 antisite defects on average are expected in each Fe-O atom column; the HAADF STEM does not have the sensitivity to resolve such small numbers of antisite defects but columns where statistically larger numbers of defects are present show an increase in atom column intensity of 10-20% in the HAADF image and intensity in the Y elemental map, confirming the presence of Y in the Fe sites. Soft x-ray absorption spectroscopy (XAS) analysis at both the Fe *L*-edge and O *K*-edge shows that Fe primarily adopts a +3 valence state (Supplementary Note 2). Given that both Fe and Y exist as trivalent ions, deviation from the ideal cation stoichiometry can be accommodated without changes in the oxygen content.

Bulk YFO is an antiferromagnet below its Néel temperature of 645 K with a net magnetization parallel to the $c_o$-axis as a result of the canting of the $Fe^{3+}$ moments[35]. Magnetometry analysis of the YFO/NSTO ($α=1.19$) shows a hysteretic response along the out-of-plane direction with a saturation magnetization ($M_s$) of ~0.056 $μ_B$/f.u., comparable with that of bulk[36], but negligible in-plane hysteresis (Fig. 2a), consistent with the magnetic anisotropy of single crystal YFO[37], and indicating that the out-of-plane orientation is the orthorhombic $c_o$-axis.

Contrary to the nonferroelectric nature of bulk, a sizeable ferroelectric polarization is observed at room temperature in the YFO/NSTO ($α=1.19$) as revealed by the polarization-electric-field (*P-E*) hysteresis loop (Fig. 2b). Positive-up-negative-down (PUND) measurement further confirms a remanent polarization ($P_r$) of 7.2 $μC/cm^2$ (Supplementary Fig. 3a). The polarization observed here is orders of magnitude stronger than the values reported previously in bulk $RFeO_3$, $R$=Gd[20], Dy[21], Ho[22] and Sm[23], and comparable with a measurement reported by Shang *et al.* in thin film YFO[28].

The leakage current is of order 1 mA/cm$^2$ at an electric field of 300 kV/cm, showing an insulating behavior.

The ferroelectric nature is confirmed by switching spectroscopy piezoresponse force microscopy (SS-PFM). The vertical PFM phase hysteresis loop, the "butterfly-shape" amplitude curve (Fig. 2c) and the piezoresponse hysteresis (Supplementary Fig. 3b) demonstrate clear reversal. This switching behavior does not qualitatively depend on the film thickness within the range of 10-100 nm, indicating that is it not greatly affected by strain relaxation, and persists at elevated temperature up to at least 150 ˚C, the limit of the instrument (Supplementary Note 4 and Figure 4). Box-in-box writing with DC voltages of +/−8 V as shown in Fig. 2d results in 180° phase contrast in the vertical PFM phase image (Fig. 2f), indicating a complete switching of polarization between upward and downward orientations. Scanning kelvin probe force microscopy (SKPFM) was performed to exclude the possibility that the PFM signal originates from non-ferroelectric mechanisms (Supplementary Note 3).

YFO was grown on SrRuO$_3$ (SRO, ~10 nm thick)-buffered STO substrates. It adopts a similar crystal structure and Y-rich stoichiometry as YFO/NSTO (Supplementary Note 5). The same ferroelectric switching behaviors are found whether the cantilever in PFM is placed directly on the YFO film surface or on the Pt top electrode (Fig. 2h-j). We also prepared YFO films on different substrates with conductive layers (see Methods), including SRO-buffered DyScO$_3$ (DSO, $a_p$=3.944 Å), La$_{0.67}$Sr$_{0.33}$MnO$_3$ (LSMO)-buffered (LaAlO$_3$)$_{0.3}$-(SrAl$_{0.5}$Ta$_{0.5}$O$_3$)$_{0.7}$ (LSAT, $a_p$=3.868 Å), and LSMO-buffered LaAlO$_3$ (LAO, $a_p$=3.788 Å). Again, Y-rich stoichiometry and ferroelectric switching are consistently observed across these samples (Supplementary Note 6), but a small dependence of the remanent piezoresponse on the substrate lattice parameter (Supplementary Table 2) suggests that the epitaxial strain plays a minor role in the piezoresponse.

*Dependence of ferroelectricity on the cation stoichiometry*

We evaluate the effects of cation stoichiometry on the ferroelectricity by growing a sample using a $Y_3Fe_5O_{12}$ (yttrium iron garnet, YIG) target first. The as-prepared "YIG" thin film on NSTO substrate shows almost the same Y/Fe ratio, *i.e.* 0.60, as the target (Supplementary Note 7), but exhibits a perovskite structure (Fig. 3a), which is qualitatively different from the garnet structure of bulk YIG. A series of $Y_\alpha FeO_{1.5(\alpha+1)}$ (0.60 < $\alpha$ < 1.19) thin films were then prepared by codeposition from the YFO and YIG targets, all of which adopt the perovskite structure (Supplementary Fig. 7b).

The ferroelectric behavior depends strongly on the Y/Fe ratio as revealed by SS-PFM. For the Y-deficient regime no switching is found (Fig. 3b, c), while for the Y-rich regime, *i.e.* $\alpha$ = 1.03 (Fig. 3d), 1.11 (Fig. 3e) and 1.19 (Fig. 2c), consistent ferroelectric switching is observed. Note that the out-of-plane lattice dimension gradually expands (Fig. 3a) as $\alpha$ becomes either greater or smaller than 1 as a consequence of cation defects. Comparing samples with $\alpha$ = 1.11 and $\alpha$ = 0.60, the lattice parameters and unit cell volume are almost the same but the ferroelectric response is markedly different, pointing to composition as the key factor. Moreover, as $\alpha$ decreases, the remanent piezoresponse gradually weakens (Supplementary Note 7). We therefore conclude that the Y/Fe stoichiometry plays the dominant role in the room temperature ferroelectricity of epitaxial YFO thin films.

*Mechanisms of ferroelectricity induced by $Y_{Fe}$ defects*

To explain the physical basis of these experimental results, first-principles DFT simulations were employed to calculate the electronic structure and atomic positions of Y-rich YFO (see Methods). The formation energy of various types of point defects in the perovskite YFO lattice was first evaluated, showing that the $Y_{Fe}$ defect has a negative formation energy for a range of in-plane strain (Supplementary Table 3), consistent with its presence in the as-prepared YFO film on various substrates. The Y-rich YFO remains insulating even at a high concentration of $Y_{Fe}$ with a band gap over 3 eV, accounting for the low leakage current.

The symmetry calculation suggests the origin of the ferroelectricity lies in the structural distortion around the $Y_{Fe}$ sites, where the *Pbnm* structure (Fig. 4a) is locally distorted into a region with *R3c*-like symmetry (Fig. 4b). In YFO the changes in bonding of oxygen atoms adjacent to $Y_{Fe}$ break the inversion symmetry, generating a net local dipole moment (Fig. 4b), which further polarizes the surrounding stoichiometric regions, producing a spontaneous polarization. The layer-resolved polarization for Y-rich YFO, calculated using the Born charge approximation, is given in Fig. 4c. Both $Y_{Fe}$ defects and epitaxial strain promote the formation of the *R3c* regions within the *Pbnm* YFO structure (Supplementary Note 8).

Position-averaged convergent beam electron diffraction (PACBED), which is sensitive to inversion center symmetry breaking[38], was then used to confirm the crystallographic origins of ferroelectricity. Compared to the simulated patterns using the DFT relaxed structures of both centrosymmetric *Pbnm* (Fig. 4e) and non-centrosymmetric *R3c* (Fig. 4f), the experimental pattern (Fig. 4g) measured from the Y-rich YFO ($\alpha$=1.19) film exhibits an asymmetric feature, *i.e.* a mismatch of intensity profiles between $[0\ k\ k]$ and $[0\ \bar{k}\ \bar{k}]$ directions (Fig. 4h), which is consistent with the simulated non-centrosymmetric *R3c*, but different from the centrosymmetric *Pbnm* pattern where the intensity profile along $[0\ k\ k]$ coincides exactly with that along $[0\ \bar{k}\ \bar{k}]$. These results support our DFT conclusions and provide evidence that the $Y_{Fe}$ defects induce inversion symmetry breaking that can support ferroelectricity.

The DFT calculations predict a switchable out-of-plane polarization of 7.2 μC/cm$^2$ for YFO with $\alpha$ = 1.28 (Supplementary Note 9), in agreement with our experimental results. The calculations also indicate that the ordering of $Y_{Fe}$ antisites has little effect on the spontaneous polarization (Supplementary Note 9). *R3c* structures in many other $R$FeO$_3$ orthoferrites with magnetic and nonmagnetic A-site ions ($R$=Lu, Yb, Er, Y, Dy, Gd, Nb, La) were also modeled. The $R_{Fe}$ antisite defects favor ferroelectric behavior (Fig. 4d) across the series, suggesting the general nature of this mechanism to yield multiferroicity in orthoferrites. In the *R3c* structure the spontaneous polarization falls monotonically with a decrease of the ionic radius of the rare earth cation.

*Integration of YFO into magnetoelectric nanocomposites*

In parallel with the understanding of the mechanism of ferroelectricity in YFO film, we also demonstrate the integration of YFO together with magnetoelastic CFO spinel into a self-assembled 2-phase nanocomposite. XRD (Fig. 5a) and cross-sectional STEM EDS mapping (Fig. 5b) reveal the formation of both perovskite and spinel phases upon codeposition of YFO and CFO. The nanocomposite consists of pillars of CFO embedded in YFO, as sketched in the inset of Fig. 5a, exhibiting coherent vertical interfaces between YFO and CFO (Fig. 5c), comparable to the self-assembled growth mechanism of BFO-CFO nanocomposites[39]. The YFO-CFO nanocomposite exhibits both robust polarization and magnetization at room temperature. The polarization originating from the YFO component (Fig. 5d) is again correlated with the presence of $Y_{Fe}$ defects seen by STEM EDS mapping (Supplementary Fig. 10a), while the magnetization (Fig. 5e) arises primarily from the CFO spinel. The saturation magnetization of ~150 emu/cm$^3$ (normalized by the CFO volume fraction) is lower than that of bulk CFO (~400 emu/cm$^3$), which is attributed to deviation from the ideal site occupancy of Co and Fe ions in inverse spinel[40] (Supplementary Note 10).

Thanks to the coherent vertical interfaces between YFO and CFO, a strain-mediated magnetoelectric coupling is expected. With an in-plane magnetic field (2000 Oe) applied *in situ*, the SS-PFM amplitude curve becomes more asymmetric: the positive coercive voltage slightly decreases while the negative coercive voltage increases, and the amplitude for a positive electric field is noticeably larger than that for a negative electric field. The clear shift of the SS-PFM phase loop toward negative electric field and the enhanced piezoresponse particularly at positive electric field (Supplementary Fig. 10c, d) collectively indicate a magnetoelectric response: an internal electric field is generated in the YFO due to strain transferred from the magnetoelastic CFO spinel upon application of the external magnetic field.

**Conclusion**

Our study reveals a mechanism for ferroelectricity based on antisite defects in the perovskite-structured canted antiferromagnet $YFeO_3$ with excess Y. The $Y_{Fe}$ defects play a dominant role in promoting a non-centrosymmetric structural distortion that yields a switchable polarization, which is robust above room temperature and present in films grown with a range of strain states and film thicknesses. The antisite defects are stable and occur without changes in oxygen content because the Y and Fe cations are isovalent, leading to a highly insulating material. Furthermore, YFO can be codeposited with a magnetoelastic spinel, CFO, forming a self-assembled magnetoelectric nanocomposite, further extending the range of functional heterostructures and nanocomposites.

According to first principles calculations, the relationship between cation off-stoichiometry and ferroelectricity is expected to occur generally within the family of orthoferrites, with magnitude depending on the specific A-site cation. Given the rich magnetic phase diagram particularly for those $R$FeO$_3$ where $R$ has partially filled $f$ electrons, it is of substantial interest and significance to investigate the influence of antisite defects on the magnetic properties and the interactions between ferroelectric and magnetic orderings. Our work therefore opens a route for promoting ferroelectricity and multiferroicity in single-phase orthoferrites.

**Methods**

**Thin film preparation:** The YFO thin films were prepared by pulsed laser deposition using a KrF excimer laser ($\lambda$=248 nm) with 1.3 J/cm$^2$ fluence and 10 Hz repetition rate to ablate a ceramic YFeO$_3$ target. The growth temperature was 900 °C and the oxygen partial pressure, $p(O_2)$, was 10 mTorr. After growth, the films were cooled down to room temperature in the same $p(O_2)$ at a rate of 20 °C/min. For those grown on nonconductive substrates, a 10-nm-thick SRO layer was grown on STO or DSO at 850 °C under $p(O_2)$=5 mTorr, and a 15-nm-thick LSMO layer was grown on LSAT or LAO at 800 °C under $p(O_2)$=10 mTorr. The "YIG" thin films were grown on NSTO substrates at the same conditions by ablating a ceramic YIG target. A series of Y$_\alpha$FeO$_{1.5(\alpha+1)}$ thin films were prepared by codeposition using YFO and YIG targets. The YFO-CFO composite films were prepared by alternately ablating YFO and CFO targets for 200 and 50 shots respectively so that each "layer" deposited from the targets is less than a monolayer thick.

**Composition and structural characterization:** The chemical composition is analyzed by using a Thermo Scientific K-Alpha+ XPS system with Al $K\alpha$ (1486.6 eV) as the X-ray source. Before collecting the data, the sample surface was cleaned with a cluster Ar ion beam for 30 seconds. For quantitative analysis of the Y/Fe ratio of YFO films, the stoichiometric YFeO$_3$ target whose Y/Fe ratio is 1:1 was measured and the ratio of the integrated areas of Y 3d and Fe 2p core level spectra was taken as a reference. XAS measurements at Fe $L$-edges and O $K$-edges were performed in total electron yield (TEY) modes using the beamline 4-ID-C of the Advanced Photon Source at Argonne National Laboratory. The temperature was set to 200 K to obtain a good signal to noise ratio. Reference scans of elemental Fe measured simultaneously indicate negligible energy shift throughout the experiments. The film thickness was analyzed by X-ray reflectivity (XRR) and the crystalline structure was characterized by both high-resolution XRD and RSM using a Rigaku SmartLab high-resolution diffractometer with Cu $K\alpha_1$ radiation (λ=1.5406 Å) as X-ray source and an incident beam Ge-(220) double-bounce monochromator.

**Transmission electron microscopy**: Cross-sectional samples of thin film YFO and nanocomposite YFO/CFO were prepared for electron microscopy by mechanical wedge polishing and final thinning using cryogenic Ar-ion milling. STEM analysis was performed with a probe corrected Thermo Fisher Scientific Titan G3 60-300 kV operated at 200 kV with a probe convergence semi-angle of 18 mrad. A collection semi-angle range of 63-200 mrad was used for STEM imaging. Atomic resolution EDS elemental maps were collected with a Thermo Fisher Scientific Super-X EDS detector and Y and Fe elemental maps were denoised using non-local principal component analysis (NLPCA) and gaussian blurring using an open-source Matlab script[41]. PACBED patterns were simulated with a custom Python-based STEM simulation using the multislice approach[42].

**Magnetic and ferroelectric preparties measurements**: Magnetic properties were analyzed using a Quantum Design MPMS-3 SQUID magnetometer or a Digital Measurement System 7035B vibrating sample magnetometer (VSM). A Precision Premier II Ferroelectric Tester was used to perform the polarization-electric field (*P-E*) loops and PUND measurements with a home-made probe station. SS-PFM, PFM and SKPFM measurements on monolithic film samples were performed on a commercial atomic force microscope (AFM) (Cypher, Asylum Research) under dual frequency resonant tracking (DART) modes with Pt-coated Si conductive probes (MikroMasch, HQ:NSC18/Pt, force constant: 2.8 N/m, tip radius: 30 nm), while *in situ* SS-PFM characterization of the nanocomposite sample was performed using another AFM (MFP-3D, Asylum Research) with a magnetic field module.

**Density function theory calculation**: First-principles calculations using density functional theory (DFT) were first performed to evaluate the formation energy of various point defects in YFO using the projector augmented wave (PAW) method as implemented in Vienna Ab-initio Simulations Package (VASP)[43,44]. The plane wave energy cutoff of 500 eV and Monkhorst-Pack k-point sampling were used. The YFeO$_3$ unit cell is calculated using an 4x4x4 k-point mesh with generalized gradient approximation (GGA) for which the Perdew–Burke–Ernzerhof (PBE) functional was used[45]. Defect calculations were performed on 80 atoms in a 2×2×1 supercell with a Monkhorst-Pack 2×2×2 mesh. The total energies and forces were converged to less than $10^{-5}$ eV and 4 meV/Å, respectively. Ferroelectric properties were calculated using the Berry-phase approach[46]. A pseudocubic ($\sqrt{2}a_o$ x $\sqrt{2}\, b_o$ x $c_o$) *Pbnm* supercell with a single Y$_{Fe}$ antisite defect and fixed in-plane lattice parameters was used to simulate Y/Fe non-stoichiometry and epitaxial growth on a SrTiO$_3$ substrate. Born effective charges approximation was used to derive atomic resolved contributions to ferroelectric polarization. To estimate polarization switching barriers we calculated the migration energy profile along the minimum energy path between two polarization states using the climbing image nudged elastic band method[47]. The choice of exchange-correlation functional is discussed in Supplementary Note 9.

**Data availability**

The data that support the findings of this study are available from the corresponding author on reasonable request.

**References**

1. Cheong, S. W. & Mostovoy, M. Multiferroics: a magnetic twist for ferroelectricity. *Nat. Mater.* **6**, 13-20 (2007).
2. Hill, N. A. Why are there so few magnetic ferroelectrics? *J. Phys. Chem. B* **104**, 6694-6709 (2000).
3. Ramesh, R. & Spaldin, N. A. Multiferroics: progress and prospects in thin films. *Nat. Mater.* **6**, 21-29 (2007).
4. Catalan, G. & Scott, J. F. Physics and applications of bismuth ferrite. *Adv. Mater.* **21**, 2463-2485 (2009).
5. Wang, J. et al. Epitaxial $BiFeO_3$ multiferroic thin film heterostructures. *Science* **299**, 1719-1722 (2003).
6. Eerenstein, W. et al. Comment on "Epitaxial $BiFeO_3$ multiferroic thin film heterostructures". *Science* **307**, 1203; author reply 1203 (2005).
7. Spaldin, N. A. & Ramesh, R. Advances in magnetoelectric multiferroics. *Nat. Mater.* **18**, 203-212 (2019).
8. Van Aken, B. B., Palstra, T. T., Filippetti, A. & Spaldin, N. A. The origin of ferroelectricity in magnetoelectric $YMnO_3$. *Nat. Mater.* **3**, 164-170 (2004).
9. Lilienblum, M. et al. Ferroelectricity in the multiferroic hexagonal manganites. *Nat. Phys.* **11**, 1070-1073 (2015).
10. Jeong, Y. K. et al. Structurally tailored hexagonal ferroelectricity and multiferroism in epitaxial $YbFeO_3$ thin-film heterostructures. *J. Am. Chem. Soc.* **134**, 1450-1453 (2012).
11. Ahn, S.-J. et al. Artificially imposed hexagonal ferroelectricity in canted antiferromagnetic $YFeO_3$ epitaxial thin films. *Mater. Chem. Phys.* **138**, 929-936 (2013).
12. Ikeda, N. et al. Ferroelectricity from iron valence ordering in the charge-frustrated system $LuFe_2O_4$. *Nature* **436**, 1136-1138 (2005).
13. Efremov, D. V., van den Brink, J. & Khomskii, D. I. Bond-versus site-centred ordering and possible ferroelectricity in manganites. *Nat. Mater.* **3**, 853-856 (2004).
14. Tokura, Y., Seki, S. & Nagaosa, N. Multiferroics of spin origin. *Rep. Prog. Phys.* **77**, 076501 (2014).
15. Fennie, C. J. & Rabe, K. M. Magnetic and electric phase control in epitaxial $EuTiO_3$ from first principles. *Phys. Rev. Lett.* **97**, 267602 (2006).

16. Lee, J. H. et al. A strong ferroelectric ferromagnet created by means of spin-lattice coupling. *Nature* **466**, 954-958 (2010).
17. Lee, J. H. & Rabe, K. M. Epitaxial-strain-induced multiferroicity in $SrMnO_3$ from first principles. *Phys. Rev. Lett.* **104**, 207204 (2010).
18. Guo, J. W. et al. Strain-induced ferroelectricity and spin-lattice coupling in $SrMnO_3$ thin films. *Phys. Rev. B* **97**, 235135 (2018).
19. Kimura, T. et al. Magnetic control of ferroelectric polarization. *Nature* **426**, 55-58 (2003).
20. Tokunaga, Y. et al. Composite domain walls in a multiferroic perovskite ferrite. *Nat. Mater.* **8**, 558-562 (2009).
21. Tokunaga, Y., Iguchi, S., Arima, T. & Tokura, Y. Magnetic-field-induced ferroelectric state in $DyFeO_3$. *Phys. Rev. Lett.* **101**, 097205 (2008).
22. Dey, K. et al. Natural ferroelectric order near ambient temperature in the orthoferrite $HoFeO_3$. *Phys. Rev. B* **100**, 214432 (2019).
23. Lee, J. H. et al. Spin-canting-induced improper ferroelectricity and spontaneous magnetization reversal in $SmFeO_3$. *Phys. Rev. Lett.* **107**, 117201 (2011).
24. Johnson, R. D., Terada, N. & Radaelli, P. G. Comment on "Spin-canting-induced improper ferroelectricity and spontaneous magnetization reversal in $SmFeO_3$". *Phys. Rev. Lett.* **108**, 219701; author reply 219702 (2012).
25. Kuo, C. Y. et al. $k = 0$ magnetic structure and absence of ferroelectricity in $SmFeO_3$. *Phys. Rev. Lett.* **113**, 217203 (2014).
26. Cheng, Z. et al. Interface strain-induced multiferroicity in a $SmFeO_3$ film. *ACS Appl Mater Interfaces* **6**, 7356-7362 (2014).
27. Shang, M. et al. The multiferroic perovskite $YFeO_3$. *Appl. Phys. Lett.* **102**, 062903 (2013).
28. Shang, M., Wang, C., Chen, Y., Sun, F. & Yuan, H. The multiferroic epitaxial thin film $YFeO_3$. *Mater. Lett.* **175**, 23-26 (2016).
29. Shang, M. et al. Multiferroicity in the $YFeO_3$ crystal. *Funct. Mater. Lett.* **13**, 1950088 (2020).
30. Haislmaier, R. C. et al. Unleashing strain induced ferroelectricity in complex oxide thin films via precise stoichiometry control. *Adv. Funct. Mater.* **26**, 7271-7279 (2016).
31. Gao, R. et al. Ferroelectricity in $Pb_{1+\delta}ZrO_3$ thin films. *Chem. Mater.* **29**, 6544-6551 (2017).
32. Lee, D. et al. Emergence of room-temperature ferroelectricity at reduced dimensions. *Science* **349**, 1314-1317 (2015).

33. Klyukin, K. & Alexandrov, V. Effect of intrinsic point defects on ferroelectric polarization behavior of SrTiO$_3$. *Phys. Rev. B* **95**, 035301 (2017).

34. Ojeda-G-P, A., Döbeli, M. & Lippert, T. Influence of plume properties on thin film composition in pulsed laser deposition. *Adv. Mater. Interfaces* **5**, 1701062 (2018).

35. White, R. L. Review of recent work on the magnetic and spectroscopic properties of the rare-earth orthoferrites. *J. Appl. Phys.* **40**, 1061-1069 (1969).

36. Jacobs, I. S., Burne, H. F. & Levinson, L. M. Field-induced spin reorientation in YFeO$_3$ and YCrO$_3$. *J. Appl. Phys.* **42**, 1631-1632 (1971).

37. Wu, A. et al. Crystal growth and magnetic property of YFeO$_3$ crystal. *Bull. Mater. Sci.* **35**, 259-263 (2012).

38. LeBeau, J. M., D'Alfonso, A. J., Wright, N. J., Allen, L. J. & Stemmer, S. Determining ferroelectric polarity at the nanoscale. *Appl. Phys. Lett.* **98**, 052904 (2011).

39. Zheng, H. et al. Self-assembled growth of BiFeO$_3$–CoFe$_2$O$_4$ nanostructures. *Adv. Mater.* **18**, 2747-2752 (2006).

40. Zhang, C. et al. Thermal conductivity in self-assembled CoFe$_2$O$_4$/BiFeO$_3$ vertical nanocomposite films. *Appl. Phys. Lett.* **113**, 223105 (2018).

41. J. Salmon, C. D., R. Willett, Z. Harmany,. Poisson noise reduction with non-local PCA. in *2012 IEEE International Conference on Acoustics, Speech and Signal Processing (ICASSP)* 1109-1112 (Kyoto, 2012).

42. Kirkland, E. J. *Advanced Computing in Electron Microscopy*, (2010).

43. Kresse, G. & Furthmuller, J. Efficient iterative schemes for *ab* initio total-energy calculations using a plane-wave basis set. *Phys. Rev. B* **54**, 11169-11186 (1996).

44. Blöchl, P. E. Projector augmented-wave method. *Phys. Rev. B* **50**, 17953-17979 (1994).

45. Perdew, J. P., Burke, K. & Ernzerhof, M. Generalized gradient approximation made simple. *Phys. Rev. Lett.* **77**, 3865-3868 (1996).

46. King-Smith, R. D. & Vanderbilt, D. Theory of polarization of crystalline solids. *Phys Rev B* **47**, 1651-1654 (1993).

47. Henkelman, G. & Jónsson, H. Improved tangent estimate in the nudged elastic band method for finding minimum energy paths and saddle points. *J. Chem. Phys.* **113**, 9978-9985 (2000).

**Acknowledgements**

The authors are grateful to Dr. Guo Tian for helpful discussions. This work was supported by the MRSEC Program of the National Science Foundation under award No. DMR-1419807 and in part by SMART, and nCORE Center of the semiconductor Research Corporation. This work was performed in part at the Center for Nanoscale Systems (CNS), a member of the National Nanotechnology Infrastructure Network (NNIN) and part of Harvard University, supported by the National Science Foundation under award No. ECS-0335765. This work also used resources of the Advanced Photon Source, a U.S. Department of Energy (DOE) Office of Science User Facility operated for the DOE Office of Science by Argonne National Laboratory under contact No. DE-AC02-06CH11357. The DFT calculations were carried out using the Extreme Science and Engineering Discovery Environment (XSEDE) which is supported by National Science Foundation Grant No. ACI-4321548562. Computational work was also partially supported by National Institute of Supercomputing and Networking/Korea Institute of Science and Technology Information with supercomputing resources including technical support (Grant No. KSC-2019-CRE-0044). A.K. acknowledges support from MIT MathWorks Engineering Fellowships. The NVIDIA Titan Xp GPU used for this research was donated by the NVIDIA Corporation.

**Author contributions**

S.N. conceived the original idea and C.A.R. supervised the project. S.N. prepared the samples and characterized the structure, composition, magnetic and ferroelectric properties with assistance from T.S. A.K. performed the TEM measurements and PACBED analysis under supervision of J.M.L. K.K., J.H.K., H.-S. K. and B.Y. performed the DFT calculations. S.N. and C.A.R. wrote the manuscript. All the co-authors discussed the results and helped to revise the manuscript.

**Competing financial interests**

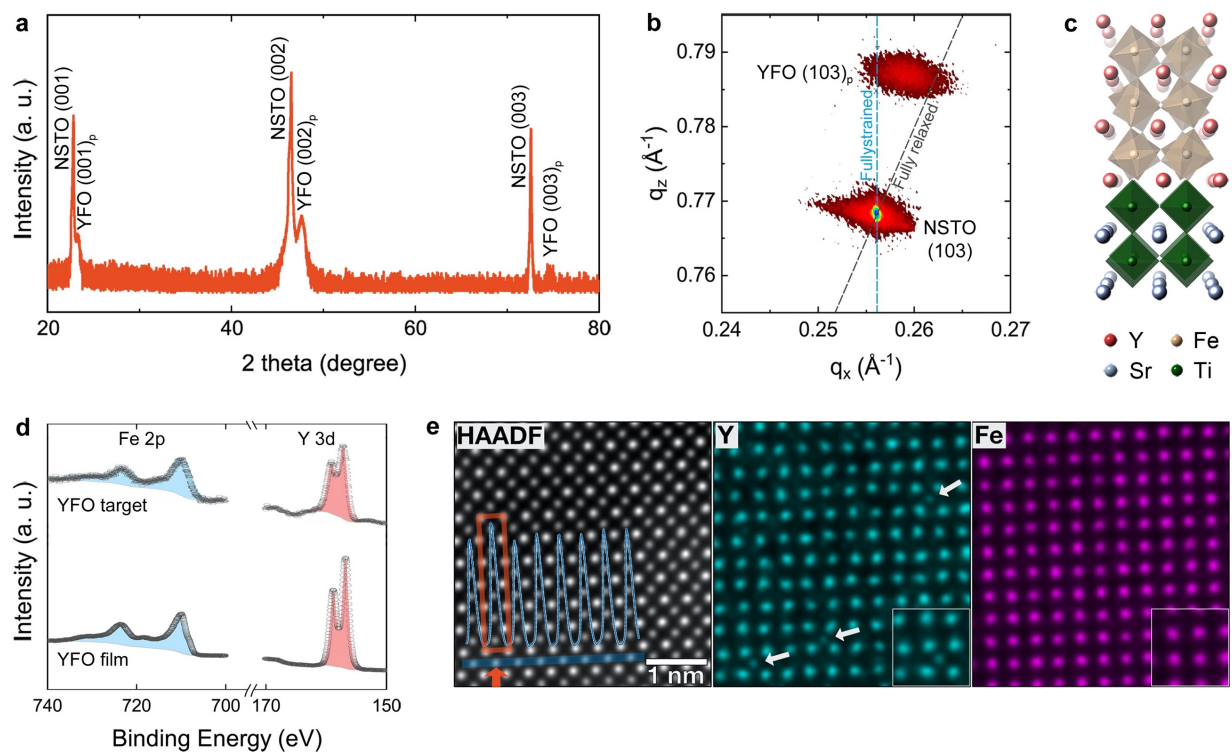

**Fig. 1 | Structure and defect analysis of YFO thin films on NSTO substrate. a**, **b**, Structural characterization of the as-grown YFO/NSTO by high-resolution XRD (**a**) and asymmetric RSM (**b**). The subscript p denotes the pseudocubic unit cell. **c**, Schematic of lattice structures of YFO and NSTO viewed along the orthorhombic [110] axis (pseudocubic [100] axis) of the YFeO$_3$ unit cell. **d**, High-resolution Fe 2p and Y 3d core level spectra of the stoichiometric YFO target and the as-prepared YFO film. **e**, HAADF STEM image and denoised atomic-resolution STEM EDS elemental mapping. The intensity profile superposed in the HAADF STEM image is taken along the blue line. Fe-O atom columns rich in Y$_{Fe}$ defects show increased intensity in HAADF (red arrow) and a signal in the Y map (white arrows).

The authors declare no competing financial interests.

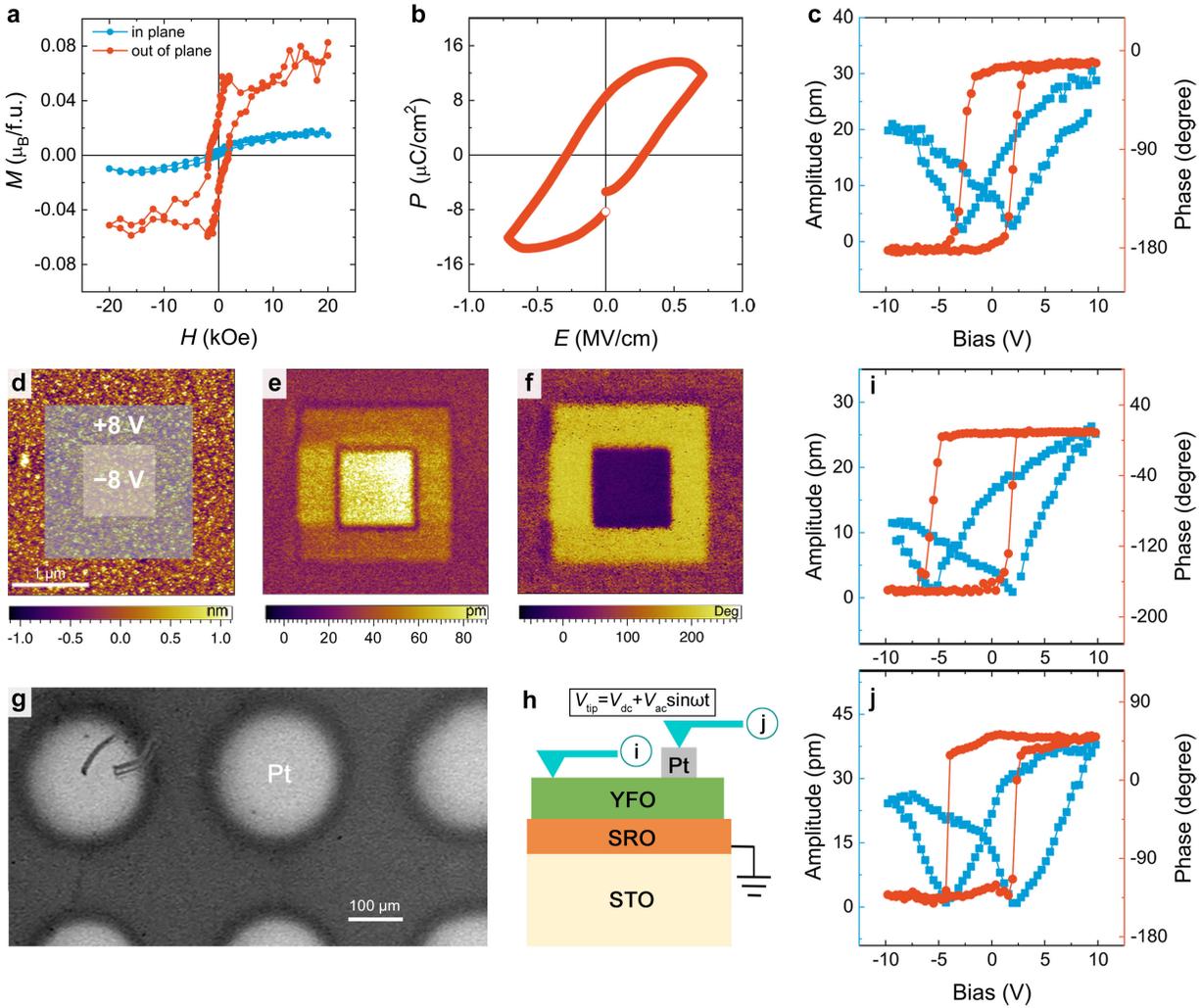

**Fig. 2 | Magnetic and ferroelectric properties of YFO thin films. a**, *M-H* curves measured at 5 K. **b**, Macroscopic *P-E* loop measured at room temperature with a frequency of 50 Hz. **c**, local SS-PFM amplitude curve and phase loop. **d**-**f**, Box-in-box writing experiments carried out as indicated in the topographic image (**d**). The vertical PFM amplitude (**e**) and vertical PFM phase (**f**) images are collected subsequently. **g**, SEM image of YFO/SRO/STO with Pt electrodes on the top. **h**, Cross-sectional schematic of SS-PFM measurements with the cantilever loaded on the surface of YFO film ot on the Pt electrode. **i**, **j**, Local SS-PFM amplitude curves and phase loops collected with the cantilever on the YFO film (**i**) and on the top of Pt electrode (**j**), respectively.

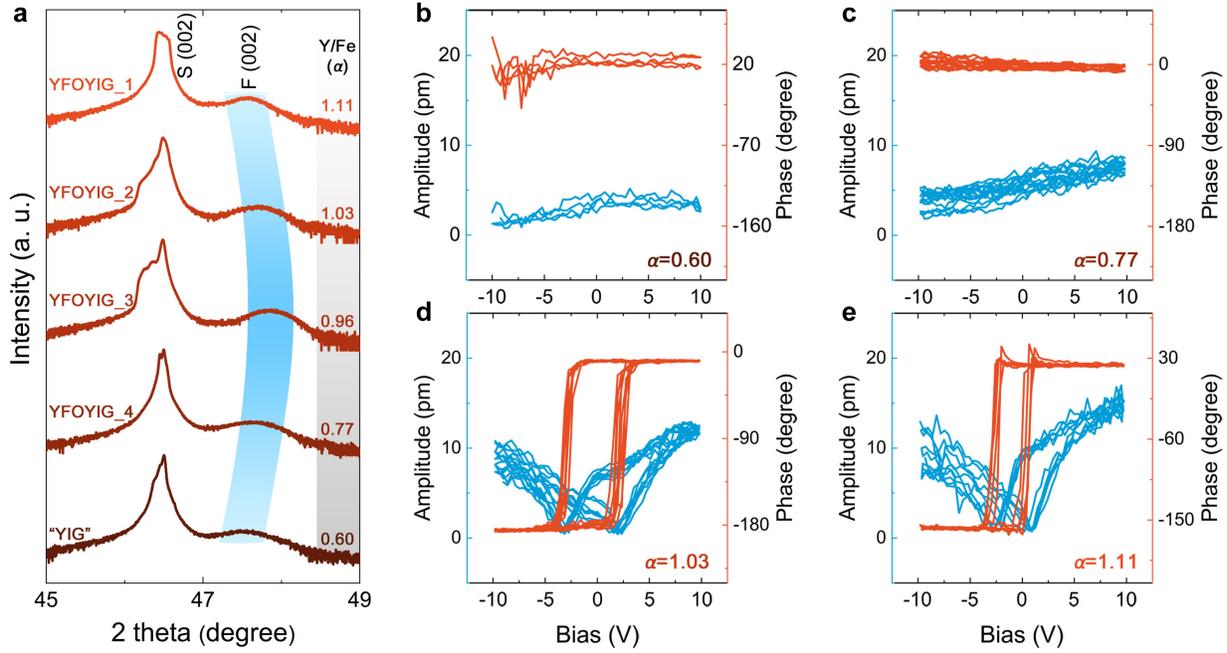

**Fig. 3 | Dependence of structure and ferroelectric switching on the Y/Fe ratio**. **a**, High-resolution XRD pattern of the 002 family of peaks of the film deposited from the YIG target and co-deposited $Y_\alpha FeO_{1.5(\alpha+1)}$ films grown on NSTO substrates. S and F refer to the peaks of substrate and film. The Y/Fe ratio, *i.e.* $\alpha$, determined by high-resolution XPS analysis is labeled correspondingly. **b-e**, Local SS-PFM amplitude curves and phase loops of "YIG" films ($\alpha = 0.60$) (**b**) and co-deposited $Y_\alpha FeO_{1.5(\alpha+1)}$ thin films with $\alpha = 0.77$ (**c**), 1.03 (**d**), and 1.11 (**e**), respectively.

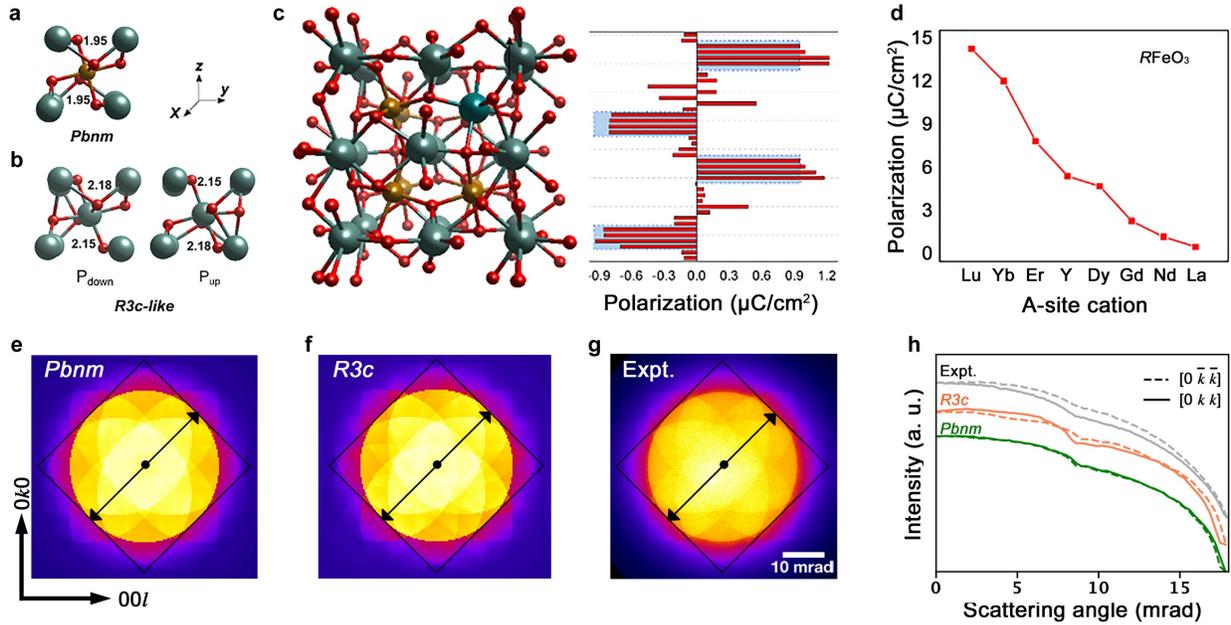

**Fig. 4 | Crystallographic origin of ferroelectricity in Y-rich YFO.** Schematics of centrosymmetric *Pbnm* (**a**) and non-centrosymmetric *R3c*-like (**b**) structures. **c**, Crystal structure and layer-resolved polarization for Y-rich YFO. Red columns indicate the atomic contribution to the ferroelectric polarization in Y-rich YFO calculated using the Born charge approximation, while blue columns show layer resolved contributions to the ferroelectric polarization in stoichiometric YFO. **d**, Ferroelectric polarization for various orthoferrites $R$FeO$_3$ that are assumed to be epitaxially grown on SrTiO$_3$ for all calculations with $R$/Fe = 1.28. **e-g**, PACBED patterns including the simulated patterns using the DFT relaxed centrosymmetric *Pbnm* (**e**) and non-centrosymmetric *R3c* (**f**) and the experimental result measured from the Y-rich YFO ($\alpha$ = 1.19) thin film sample (**g**). **h**, Intensity profiles integrated across the pattern diagonals from PACBED patterns along the black arrows shown in (**e-g**).

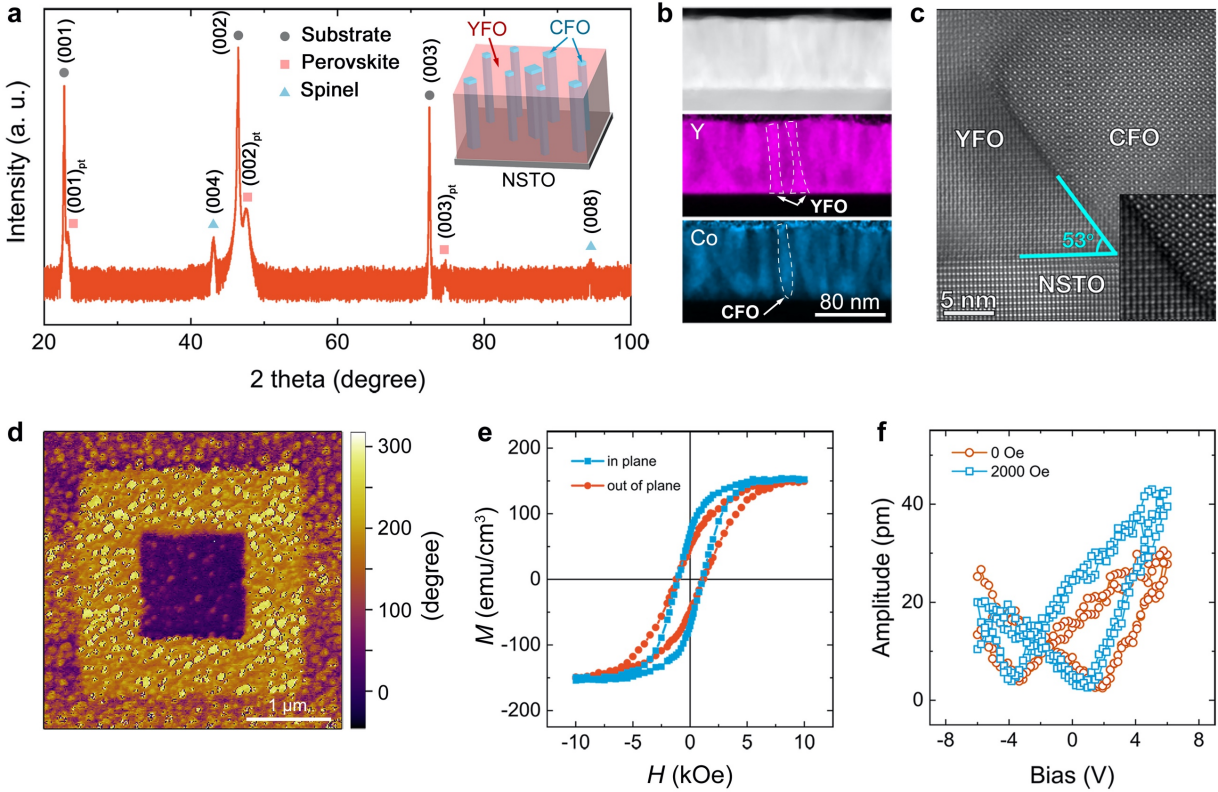

**Fig. 5 | Self-assembled multiferroic YFO-CFO nanocomposites**. **a**, High-resolution XRD pattern of the YFO-CFO composite film on NSTO substrate. The inset is a schematic of the nanocomposite. **b**, Cross-sectional TEM image and EDS elemental mapping indicate Co in the CFO pillar and Y in the YFO matrix as outlined by dashed lines. **c**, HAADF STEM image taken at the interface with the zone axis along the [110] direction of NSTO substrate, showing both epitaxial CFO and YFO. **d**, Vertical PFM phase contrast image taken after poling with voltages of +/ −8 V. The scattered features with brighter contrast correspond to the protruding CFO pillars. **e**, *M-H* curves measured at room temperature. **f**, *in situ* local SS-PFM amplitude curves measured with and without an in-plane magnetic field (2000 Oe) applied.

Supplementary Information

# An antisite defect mechanism for room temperature ferroelectricity in orthoferrites

Shuai Ning[1*], Abinash Kumar[1], Konstantin Klyukin[1], Jong Heon Kim[2], Tingyu Su[1], Hyun-Suk Kim[2], James M. LeBeau[1], Bilge Yildiz[1,3] and Caroline A. Ross[1*]

*Contents:*

**Supplementary Note 1. STEM and XPS analysis.**

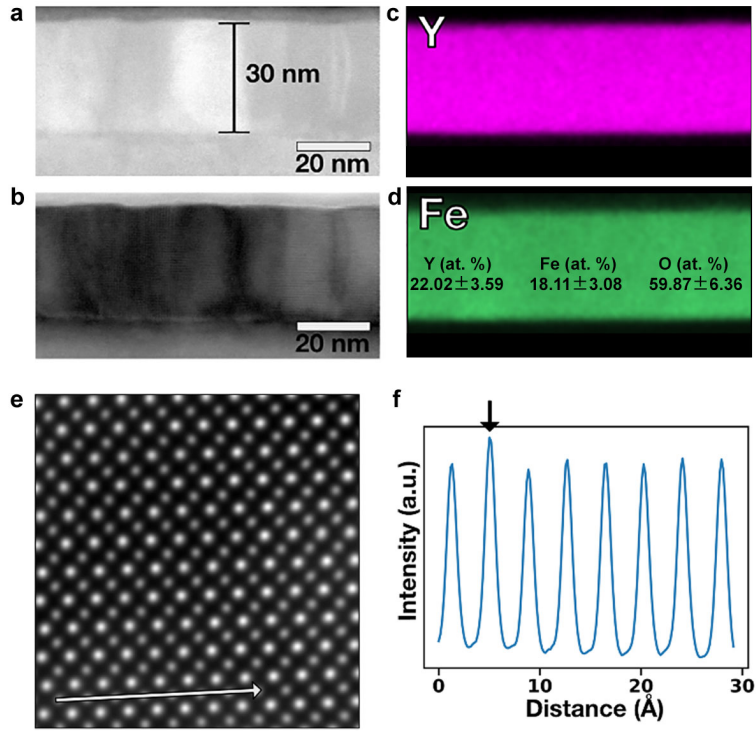

**Supplementary Fig. 1 | STEM of the YFO/NSTO sample. a-d**, Cross-sectional STEM image of high-angle annular dark-field (HAADF) (**a**) and bright field (BF) (**b**) collected from a YFO thin film grown on a NSTO substrate in cross section along the pseudocubic [100], and EDS mapping of Y (**c**) and Fe (**d**). BF STEM image reveals that the YFO thin film shows mosaicity with slight deviations of the crystal orientations. **e**, High-resolution HAADF STEM image of the film. **f**, HAADF STEM intensity profile along the line shown in (**e**). Note the Fe-O atom column with increased intensity (arrow) contains $Y_{Fe}$ defects.

**Supplementary Table 1 | Composition analysis of YFO films on NSTO substrate prepared from different PLD runs under the same conditions.**

|        | EDS  | High resolution XPS |      |      |      |      |           |
|--------|------|------|------|------|------|------|-----------|
| Sample | #1   | #1   | #2   | #3   | #4   | #5   | Average   |
| Y/Fe   | 1.21 | 1.19 | 1.22 | 1.20 | 1.23 | 1.13 | 1.19±0.04 |

**Supplementary Note 2. XAS analysis.**

The chemical states are analyzed by XAS at both the Fe *L*-edge and O *K*-edge. The Fe *L*-edge XAS spectra results from Fe 2p to Fe 3d transitions. Specifically, peak I (707.6 eV) arises from transitions to non-bonding Fe 3d $t_{2g}$ orbitals, while Peak II (709.3 eV) is ascribed to transitions to anti-bonding Fe 3d $e_g$ states. This $L_3$-edge feature confirms that Fe adopts the +3 valence state as seen in stoichiometric bulk orthoferrites [1].

Further support is provided by the O *K*-edge XAS spectra, which corresponds to the excitation of O 1s electrons to anti-bonding states that originate from O 2p orbitals interacting with orbitals on the coordinating atoms, in which A, A′, C, and C′ are assigned to transitions O 2p to Fe 3d $t_{2g}$, O 2p to Fe 3d $e_g$, and O 2p to Fe 4p/4s states respectively, while B and B′ are related to the interactions between O 2p and Y 4d orbitals. In particular, the crystal field splitting of 1.6 eV, *i.e.* the difference between A and A′, is close to the value of 1.7 eV, the difference between Peak I and II in the Fe $L_3$-edge spectra, are in good agreement with a previous study on bulk YFeO$_3$ compounds [1], indicating negligible oxygen deficiency in the YFO films.

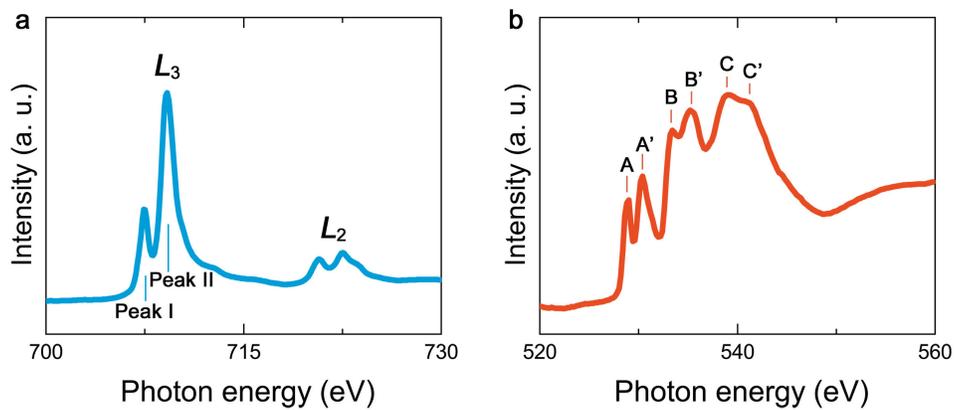

**Supplementary Fig. 2 | XAS analysis of as-prepared YFO thin films.** XAS spectra at Fe *L*-edge (**a**) and O *K*-edge (**b**).

**Supplementary Note 3. Ferroelectric property analysis on YFO/NSTO.**

PUND measurements with pulse width of 1 ms and pulse delay time of 1000 ms were performed on the YFO/NSTO sample. Polarization is measured after each pulse. The total polarization ($P^*$) was calculated at the first pulse (switching) which contains the contribution of both switched charge density ($Q_{SW}$) and leakage current. The second pulse (non-switching) measured the amount of polarization ($P^\wedge$) excluding the $Q_{SW}$. After applying these two pulses in positive direction, two negative pulses were applied. As Supplementary Fig. 3a shows, the switched charge density $Q_{SW}$, *i.e.* the remnant polarization, calculated by $P^* - P^\wedge$, is 7.2 μC/cm$^2$, slightly smaller than that in the *P-E* loop, which is a common phenomenon.

Supplementary Fig. 3b shows the piezoresponse hysteresis, PR(V), loop that is calculated according to PR(V) = Amp(V) • cos [$\phi$(V)] with the amplitude, Amp(V), and phase, $\phi$(V), shown in Fig. 2c, which clearly demonstrates the ferroelectric characteristic of the YFO film on NSTO substrate.

Supplementary Fig. 3c, d show the lateral PFM signals collected simultaneously with the vertical ones shown in Fig. 2e, f. Unlike the vertical PFM phase image (Fig. 2f), no clear 180° phase contrast was seen in the lateral PFM phase image (Supplementary Fig. 3d) between the regions poled by opposite voltages, indicating there is negligible in-plane component of the polarization. The contrast in both lateral PFM amplitude (Supplementary Fig. 3c) and lateral PFM phase (Supplementary Fig. 3d) images is only observed at the boundaries between regions written by opposite voltages can be a result of surface charge trapping.

Scanning kelvin probe force microscopy (SKPFM) was performed to rule out possible non-ferroelectric mechanisms, *e.g.* surface-trapped charges, that may contribute to signals in PFM characterization. After poling, an apparent surface potential contrast is observed (Supplementary

Fig. 3e): the positively (negatively) poled area shows higher (lower) surface potential. After multiple contact-mode scans over the same area with the tip grounded, legible surface potential contrast remains (Supplementary Fig. 3f) except for a slight decrease in the potential difference due to the removal of excessive screening charges on the surface. The remaining surface potential contrast arises from the difference in the surface charges that assist in stabilizing the polarization with opposite orientations poled by negative and positive biases.

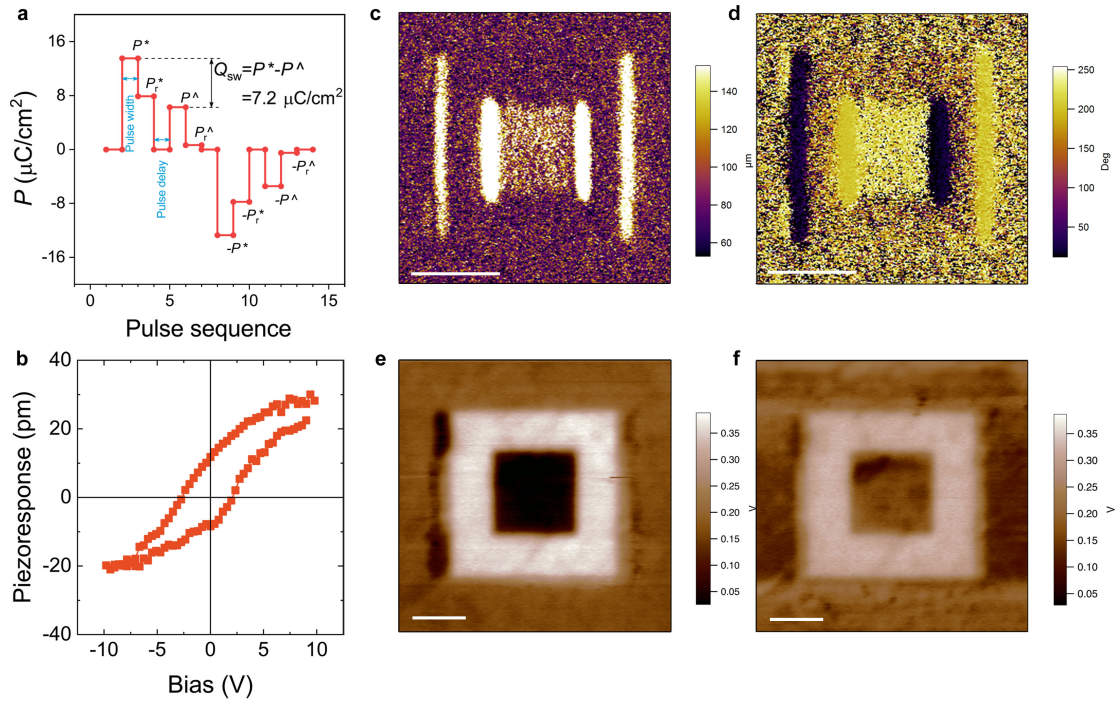

**Supplementary Fig 3 | Ferroelectric properties of YFO/NSTO. a**, PUND measurements on YFO/NSTO. The pulse width was 1 ms and the pulse delay was 100 ms. **b**, Piezoresponse hysteresis loop calculated with the amplitude and phase data. **c**, **d**, Lateral PFM amplitude (**c**) and phase (**d**) images. **e**, **f**, SKPFM images taken right after the poling (**e**) and after multiple contact-mode scans with tip grounded (**f**), respectively. The scale bars in (**c-f**) all correspond to 1μm.

**Supplementary Note 4. SS-PFM of YFO/NSTO with varying thickness and temperature.**

YFO thin films with thickness ranging from 10 to 100 nm on NSTO substrates were prepared under the same PLD conditions. SS-PFM measurements reveal that ferroelectric behavior is always present irrespective of the film thickness, as seen in Supplementary Fig. 4a, b. Note that the 10-nm-thick film will break down under bias >7 V.

The measurements were also performed at elevated temperatures for the 30-nm-thick YFO film, *i.e.* 100 °C and 150 °C (Supplementary Fig. 4c, d). The results suggest that the ferroelectric switching still persists at temperature up to 150 °C, the limit of the instrument. Compared to that measured at room temperature (Fig. 2c), we notice that coercive voltage increases significantly at elevated temperature. This is due to the fact that the ferroelectric polarization usually becomes more active as temperature increases. Higher voltages are required to fully switch and stabilize the polarization [2].

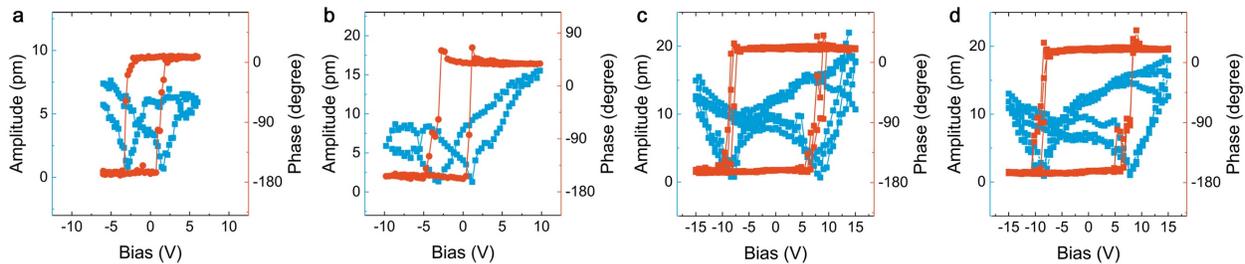

**Supplementary Fig. 4 │ SS-PFM of YFO/NSTO with varying thicknesses and temperature.** SS-PFM amplitude curves and phase loops of the YFO thin films with thickness of 10 nm (**a**) and 100 nm (**b**) measured at room temperature, and of the 30-nm-thick YFO films measured at elevated temperatures, *i.e.* 100 °C (**c**) and 150 °C (**d**), respectively.

**Supplementary Note 5. YFO thin films grown on SRO/STO.**

High-resolution XPS analysis shows that the YFO film grown on SRO/STO is also Y-rich with a Y/Fe ratio of 1.11. High-resolution XRD (Supplementary Fig. 5a) and (013) RSM (Supplementary Fig. 5b) indicate the YFO film also exhibits a slight strain relaxation, similar to that grown on NSTO substrate and attributed to imperfections like mosaicity. Almost the same piezoresponse hysteresis loops were obtained with the cantilever loaded directly on the surface of YFO thin film and on the Pt top electrode (Supplementary Fig. 5c, d). Box-in-box writing and reading were also performed on YFO/SRO/STO. Vertical PFM amplitude and phase images (Supplementary Fig. 5e, f) clearly reveal domain switching.

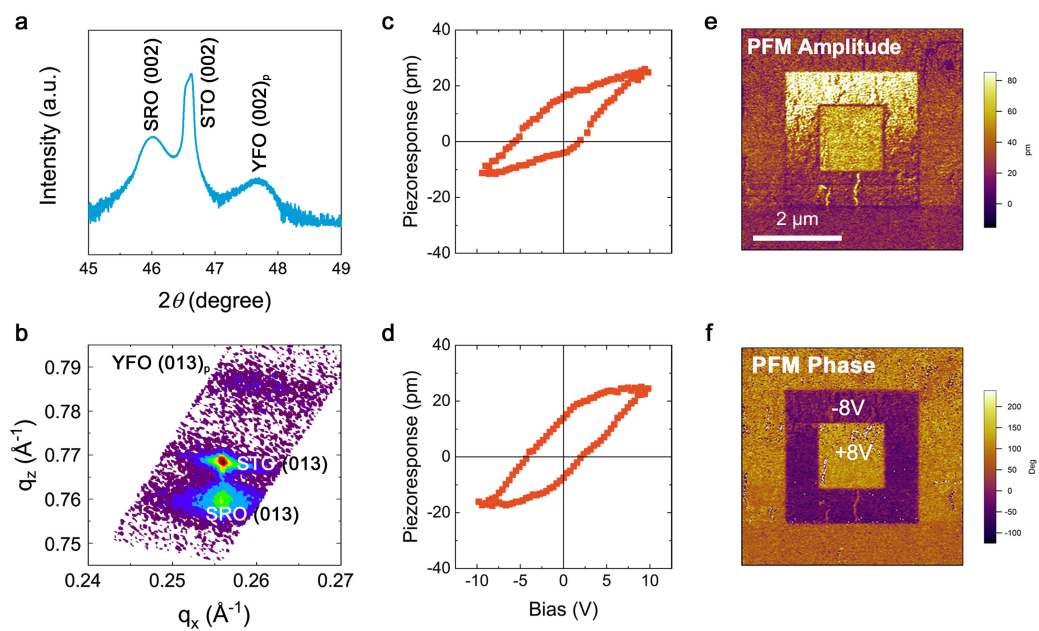

**Supplementary Fig. 5 | Structural and PFM characterization on YFO/SRO/STO. a, b,** Structure characterization of YFO/SRO/STO by XRD (**a**) and RSM (**b**). **c, d,** Calculated piezoresponse hysteresis loops with the cantilever loaded directly on the surface of YFO film (**c**) and on the top of Pt electrode (**d**). **e, f,** The vertical PFM amplitude (**e**), and vertical PFM phase (**f**) images collected after writings with voltages of +/−8 V applied on the tip.

**Supplementary Note 6. YFO thin films on different substrates.**

We prepared YFO thin films under the same conditions on different substrates with suitable conductive layers. High-resolution XPS was conducted to analyze the chemical composition, and the Y/Fe ratio is 1.12, 1.19 and 1.17 for the YFO films grown on SRO/DSO, LSMO/LSAT and LSMO/LAO (Supplementary Table 2), respectively.

Structural characterization shows that the 10~15 nm-thick conductive layers are coherently strained to the substrates in all cases, allowing us to evaluate the effects of epitaxial strain (ranging from +2.5% to −1.5%) on the ferroelectricity of the YFO films. With a decrease of the substrate lattice parameter, the out-of-plane lattice parameter ($c_p$) of YFO thin films increases indicated by the left shift of the 002 peak (Supplementary Fig. 6a). For the YFO film grown on SRO/DSO (Supplementary Fig. 6b), a significant relaxation is present, which is similar to that grown on SRO/STO (Supplementary Fig. 5b). For the YFO grown on LSMO/LSAT (Supplementary Fig. 6c), the film peak is largely overlapped with the substrate peak due to the similar lattice parameters but a slight relaxation can be noticed. For the YFO grown on LSMO/LAO, the film peak is overlapped with the Kiessig fringes of the LSMO layer (Supplementary Fig. 6a, d). While the out-of-plane lattice parameter can be extracted from the high resolution XRD, the identification of the YFO peak in RSM and hence the interpretation of in-plane lattice parameter is challenging, but if we assume the YFO is lattice matched to the substrate, the unit cell volume is smaller than bulk. Our DFT calculations indeed indicate a lattice volume shrinkage of -0.19% for a compressive strain of -1.0%, but a lattice expansion of +3.2% for a tensile strain of +1.0%. This trend has also been reported in other perovskite oxides, *e.g.* $SrRuO_3$ [3] and $LaCoO_3$ [4]. The LAO substrate with LSMO layer in our work provides a large compressive strain for YFO. Therefore, it is reasonable that the unit cell volume is smaller than that of bulk.

We summarize the lattice parameters of the YFO thin films grown on different substrates in Supplementary Table 2, with those of YFO/SRO/STO included as well. With strain varying from tensile to compressive, the c/a ratio changes from smaller to greater than 1. In addition, the pseudocubic unit cell volume shows a shrinkage for the compressive strain regime. Despite these differences in the strain states, room temperature ferroelectric switching is obtained consistently (Supplementary Fig. 6e-g). The calculated piezoresponse hysteresis loops show that with the strain varying from tensile to compressive the maximum piezoresponse gradually decreases (Supplementary Table 2).

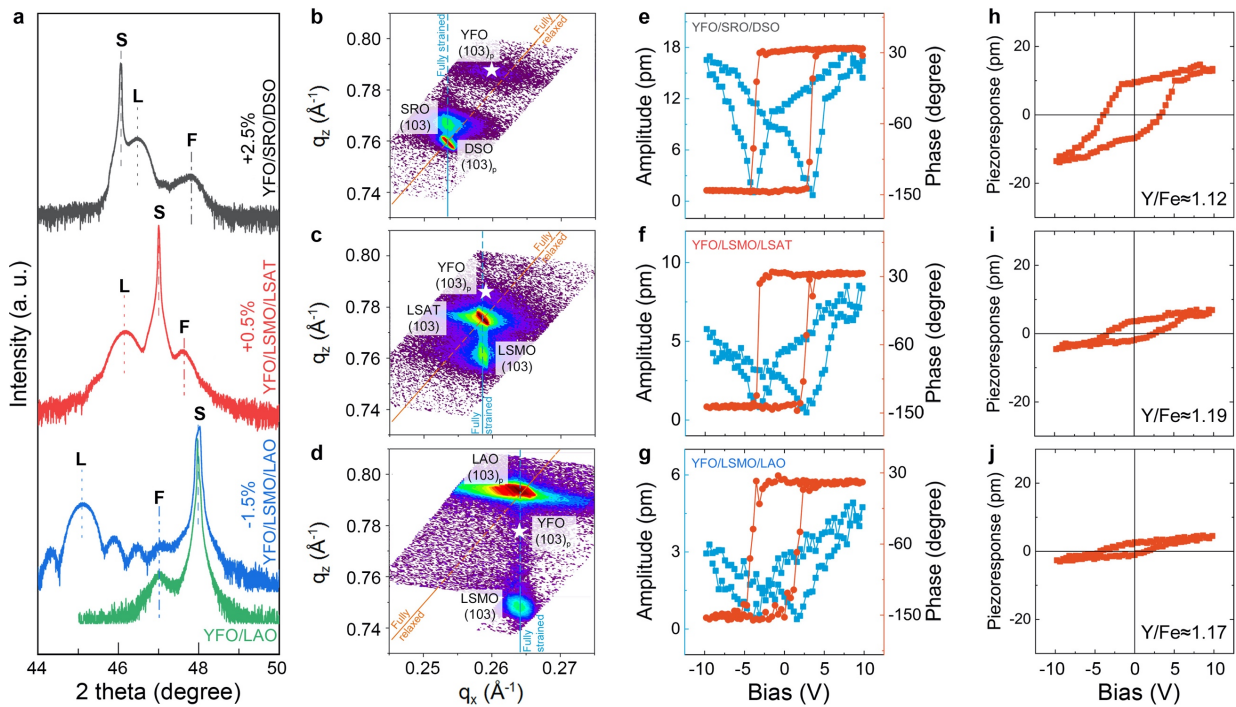

**Supplementary Fig. 6 | Structure and ferroelectric switching of YFO films on different substrates. a**, High-resolution XRD pattern of the 002 family of peaks of YFO thin films grown on SRO/DSO, LSMO/LSAT, LSMO/LAO, and bare LAO substrates. The **S**, **L** and **F** refer to the peaks of substrate, conductive layer, and YFO thin film, respectively. **b-d**, 013 RSM of YFO films grown on SRO/DSO (**b**), LSMO/LSAT (**c**) and LSMO/LAO (**d**), respectively. The blue and orange dash lines indicate fully strained and fully relaxed states, respectively. The film peaks of those grown on LSMO/LSAT and LSMO/LAO are partially overlapped with the substrate signals. **e-g**,

Local vertical SS-PFM amplitude curves and phase loops of the films grown on SRO/DSO (**e**), LSMO/LSAT (**f**) and LSMO/LAO (**g**), respectively. **h-j**, Piezoresponse hysteresis loops of films grown on SRO/DSO (**h**), LSMO/LSAT (**i**) and LSMO/LAO (**j**), respectively.

**Supplementary Table 2 │ Lattice parameters of YFO films grown on different substrates**

| Sample | Substrate lattice parameter (Å) | YFO film lattice parameters (Å) | | unit cell volume (Å$^3$) | Y/Fe ratio | 2 × remnant piezoresponse (pm) |
|---|---|---|---|---|---|---|
| | | $a_p$ | $c_p$ | | | |
| YFO/SRO/DSO | 3.944 | 3.862 | 3.812 | 56.83 | 1.12 | 18.5 |
| YFO/SRO/STO | 3.905 | 3.861 | 3.817 | 56.90 | 1.11 | 20.2 |
| YFO/LSMO/LSAT | 3.868 | 3.856 | 3.829 | 56.93 | 1.19 | 5.61 |
| YFO/LSMO/LAO | 3.788 | 3.789* | 3.865 | 55.49 | 1.17 | 3.81 |
| *Bulk YFO* | *N/A* | *3.847* | *3.803* | *56.28* | *1.00* | *N/A* |

\* This value is estimated assuming the YFO thin film is fully strained to the substrate. A strain relaxation due to the large lattice mismatch would lead to a larger in-plane lattice parameter and unit cell volume.

**Supplementary Note 7. Structure and ferroelectric switching of $Y_\alpha FeO_{1.5(\alpha+1)}$ thin films.**

The Y/Fe ratio of the "YIG" film grown by using the $Y_3Fe_5O_{12}$ target was analyzed by high-resolution XPS. Calibrated with the stoichiometric $YFeO_3$ target, the ratio of the integrated areas of Y 3d and Fe 2p core level spectra (Supplementary Fig. 7a) indicates a Y/Fe ratio of 0.60, or $Y_{0.75}Fe_{1.25}O_3$. Unexpectedly, the as-grown $Y_{0.75}Fe_{1.25}O_3$ film on NSTO substrate exhibits a perovskite structure (Supplementary Fig. 7b). To further study the morphology this Y-deficient sample, cross-sectional STEM imaging and EDS mapping were performed. This shows compositional heterogeneity with regions having excessive Y or Fe respectively (Supplementary Fig. 7d). The Y-poor (Fe-rich) region could be either due to enrichment of Fe or formation of Y vacancies. Fe antisites on Y sublattice were not found in this sample, ruling out the possibility of Fe enrichment. Formation of Y vacancies is likely because much weaker EDS intensity can be observed as the circle indicates in Supplementary Fig. 7f. Despite the heterogeneity in composition, both Y-poor and Y-rich regions exhibit a perovskite structure (Supplementary Fig. 7e).

To further characterize the structure, a high resolution reciprocal space map around the 013 family of peaks was collected. As shown in Supplementary Fig. 7c, besides the 013 peak of NSTO substrate, only one broad film peak is present, indicating no apparent structural phase separation.

The codeposited $Y_\alpha FeO_{1.5(\alpha+1)}$ thin films all adopt the perovskite structure (Supplementary Fig. 7b), but the lattice parameter varies with $\alpha$, indicated by the slight shift of film peaks. In other words, the lattice will gradually expand upon the incorporation of defects in both the Y-rich and the Y-deficient cases.

The piezoresponse hysteresis loops of samples with $\alpha = 1.11$ and $\alpha = 1.03$ are shown in Supplementary Fig. 7g, h. Plotting the 2 × remanent piezoresponse as a function of Y/Fe ratio ($\alpha$) in Supplementary Fig. 7i, the piezoresponse increases as the sample becomes more Y-rich.

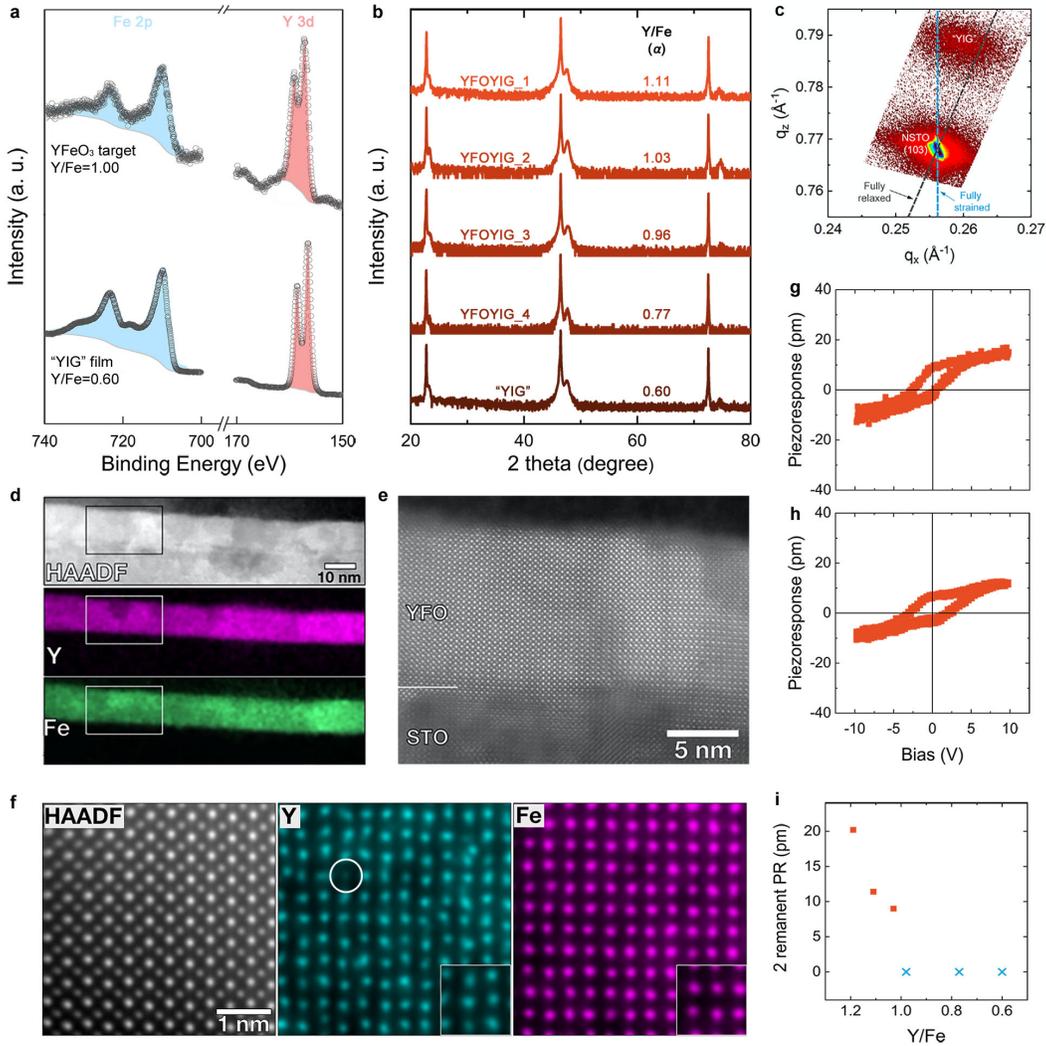

**Supplementary Fig. 7 | "YIG" ($Y_{0.75}Fe_{1.25}O_3$) and $Y_\alpha FeO_{1.5(\alpha+1)}$ thin films**. **a**, High-resolution XPS of Fe 2p and Y 3d core level spectra of the stoichiometric $YFeO_3$ target and the "YIG" $Y_{0.75}Fe_{1.25}O_3$ film on NSTO substrate. **b**, XRD patterns of the $Y_{0.75}Fe_{1.25}O_3$ film and codeposited $Y_\alpha FeO_{1.5(\alpha+1)}$ thin films. **c**, 013 RSM of the $Y_{0.75}Fe_{1.25}O_3$ film. **d**, HAADF STEM image and EDS maps of $Y_{0.75}Fe_{1.25}O_3$ film on NSTO substrate. **e**, Enlarged HAADF STEM image including both Y-poor and Y-rich regions as indicated by the rectangular in (**d**). **f**, Atomic-resolution STEM EDS elemental mapping on the Y-poor region. The circle reveals a signature of Y deficiency due to its much weaker intensity. **g**, **h**, Calculated piezoresponse hysteresis loops of samples with $\alpha = 1.11$ (**g**) and $\alpha = 1.03$ (**h**). **i**, 2 × remnant piezoresponse as a function of Y/Fe ratio ($\alpha$).

**Supplementary Note 8. DFT calculation of YFeO$_3$**

First-principles calculations using density functional theory (DFT) were performed to evaluate the defect formation energy of various types of point defects. YFO structures under different strain states, *i.e.*, bulk (unstrained), and with compressive (-1.0%) and tensile (+1.0%) in-plane strains were considered. The defect formation energy was calculated, which is defined as $E_{form}=E_{defect}-E_{perfect}+\sum n_i\mu_i$, where $E_{defect}$ is the total energy of the supercell with the defect, $E_{perfect}$ is the total energy of the perfect intrinsic supercell, $n_i$ denotes the number of atoms, and $\mu_i$ represents the chemical potential of the corresponding atom. All point defects were considered in their neutral states. As Supplementary Table 3 shows, for all the strain states considered, the Y$_{Fe}$ defect has a negative formation energy, contrary to other types of point defect, indicating that creating YFO with excess Y accommodated by Y$_{Fe}$ defect is more energetically favorable than growing stoichiometric YFO.

The calculated electronic properties show that Y-rich YFO ($\alpha$=1.28) remains insulating even at high antisite defect concentration (Supplementary Fig. 8a). We then compared the relative stability of different YFeO$_3$ phases: centrosymmetric *Pbnm* and two ferroelectric phases (*Pna21* and *R3c*) under different epitaxial strain and Y/Fe stoichiometry. *Pna21* phase was found to be unstable for all considered conditions, while *R3c* symmetry can be stabilized by a large epitaxial strain and Y/Fe non-ideal stoichiometry as seen in Supplementary Fig. 8b. Although this phase remains metastable under experimental conditions (*e.g.* grown on DSO, *a* =3.944 Å and STO, *a* =3.905 Å) substrates), the presence of Y$_{Fe}$ antisite defects could stabilize *R3c*-like regions within the *Pbnm* structure. Note that the epitaxial strain is essential for the stabilization of *R3c*-like regions which are not observed in fully relaxed supercells.

**Supplementary Table 3 | Formation energy of point defects in YFO structure under different strain states.**

|     | Bulk | Compressive (-1.0%) | Tensile (+1.0%) |
| --- | --- | --- | --- |
| $V_O$ | 3.271 eV | 4.121 eV | 5.032 eV |
| $V_{Fe}$ | 1.856 eV | 1.686 eV | 3.034 eV |
| $V_Y$ | 6.887 eV | 6.444 eV | 8.234 eV |
| $Fe_i$ | 5.664 eV | 6.054 eV | 4.623 eV |
| $Y_i$ | 3.564 eV | 3.929 eV | 3.121 eV |
| $Fe_Y$ | 6.890 eV | 8.259 eV | 6.799 eV |
| $Y_{Fe}$ | **-1.746 eV** | **-1.704 eV** | **-0.742 eV** |

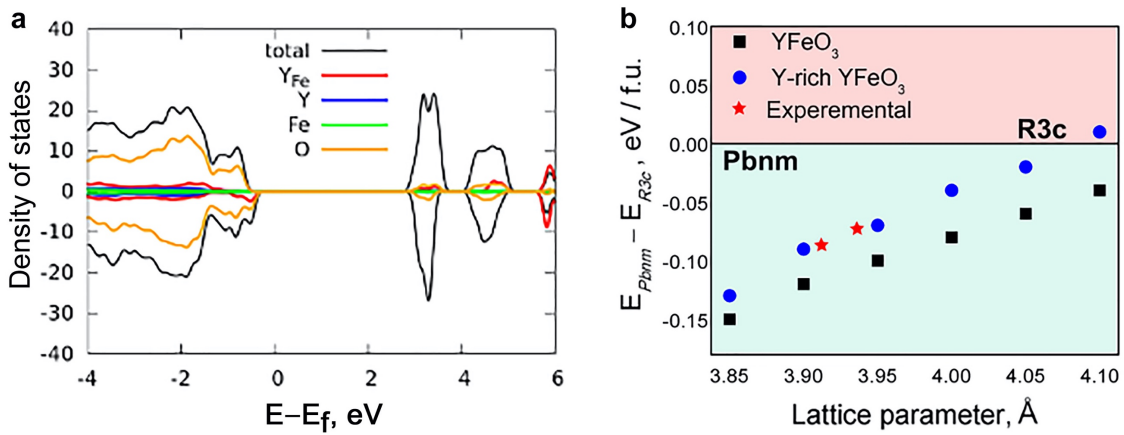

**Supplementary Fig. 8 | Electronic structure and relative stability of *Pbnm* and *R3c* phases.** **a**, projected density of states of Y-rich YFO ($\alpha = 1.28$). Fermi level is set to zero. **b**, the relative stability of *Pbnm* and *R3c* phases of pristine YFO and Y-rich YFO ($\alpha = 1.28$). Stars indicate substrate lattice parameters considered in our experiment.

**Supplementary Note 9. Choice of exchange-correlation functional to describe ferroelectricity.**

The choice of exchange-correlation functional is known to greatly affect spontaneous polarization and switching barriers in ferroelectric materials [5]. We investigated the effect of various exchange-correlation functional and their combination with a small Coulomb U on ferroelectric polarization and switching barrier of Y-rich $YFeO_3$.

Our calculations revealed that ferroelectric properties of Y-rich $YFeO_3$ decrease significantly when DFT+U with large effective Coulomb U ($U_{eff}$ =4 eV) are employed (Supplementary Table 4). The spontaneous polarization of Y-rich $YFeO_3$ calculated using LDA+U ($U_{eff}$=1 eV), PBE, PBEsol and PBEsol+U ($U_{eff}$ =1 eV) and HSE functional lies in the range of 3.2-7.2 $\mu C/cm^2$, in a reasonable agreement with experimental results. Our calculations also indicate that the polarizations of neighboring $Y_{Fe}$ do not cancel out each other and almost equal polarization amplitude is observed for both ordered and arbitrarily distributed $Y_{Fe}$ antisites at Y/Fe = 1.13 (Supplementary Fig. 9).

The switching barrier calculated with the discussed above methods are underestimated with the largest activation energy of 57 meV calculated using LDA+U ($U_{eff}$ =1 eV) method. The latter approach is chosen to calculate ferroelectric properties of other orthoferrites (Figure 4d in the main text). HSE hybrid functional was used to reduce a self-interaction error in the electronic structure calculations.

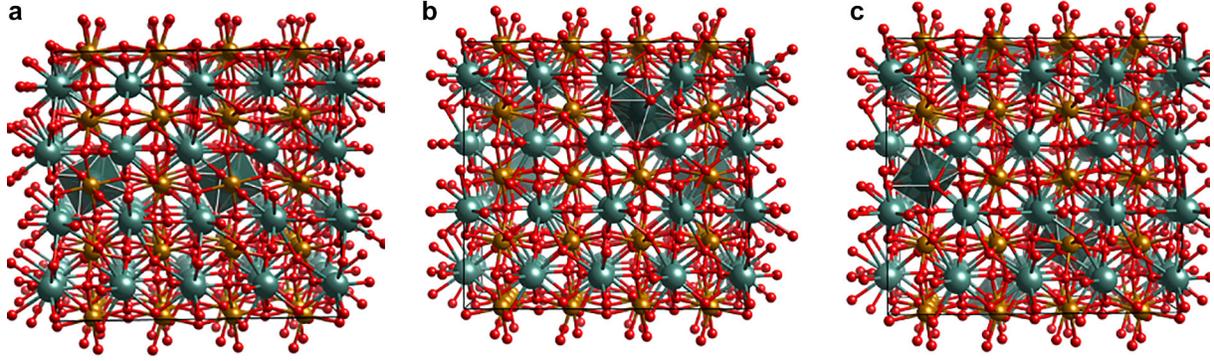

**Supplementary Fig. 9 |** Calculated ferroelectric polarization for three different $Y_{Fe}$ defect distributions (Y/Fe = 1.13) in the YFO epitaxially grown on SrTiO$_3$ substrate. Y, Fe and O atoms are in green, brown and red. $Y_{Fe}$ positions are highlighted with polyhedra. The calculated polarization along (001) direction is 3.4, 2.9 and 3.6 µC/cm$^2$ for YFO with ordered (a), cluster-like (b) and arbitrarily distributed (c) $Y_{Fe}$ defects, respectively.

**Supplementary Table 4 |** Ferroelectric properties of Y-rich YFeO$_3$ using LDA, PBE, PBEsol functional and their combination with Coulomb U.

| Functional | Polarization, µC/cm$^2$ | Switching barrier, meV |
|---|---|---|
| PBE | 6.6 | 43 |
| LDA | metallic | - |
| PBEsol | 6.2 | 27 |
| PBEsol+U1 | 4.2 | 21 |
| PBEsol+U4 | 0.9 | 4 |
| LDA+U1 | 7.2 | 57 |
| LDA+U4 | 1.3 | 5 |
| HSE | 3.2 | 18 |

* Calculations are carried out for $\sqrt{2}a_p \times \sqrt{2}b_p \times c_p$ *Pbnm* supercell assuming epitaxial growth on STO substrate. Ferroelectric polarization along out-of-plane ($c_p$) direction is calculated using Berry Phase approach. Switching barriers are calculated using CI-NEB approach.

**Supplementary Note 10. Self-assembled YFO-CFO nanocomposite film.**

HAADF-STEM and EDS mapping collected on YFO components of the YFO-CFO nanocomposite film (Supplementary Fig. 10a) shows the presence of $Y_{Fe}$ defects indicated by the arrows. In Supplementary Fig. 10b, the schematic of the CFO lattice is superposed on the STEM EDS elemental mapping of CFO component, in which the cyan balls represent the ions at tetrahedral sites while the red balls correspond to the ions at octahedral sites.

The chemical formula unit of stoichiometric cobalt ferrites can be written as $[Co^{2+}_{1-x}Fe^{3+}_{x}]_{tet}$ $[Co^{2+}_{x}Fe^{3+}_{2-x}]_{oct}O_4$ where the x is the inversion parameter defined as the fraction of tetrahedral sites occupied by $Fe^{3+}$. The magnetic moments on the tetrahedral and octahedral sites are coupled antiferromagnetically, so for the stoichiometric cobalt ferrite that is completely inverse with x=1 and can be written as $[Fe^{3+}]_{tet}[Co^{2+}\ Fe^{3+}]_{oct}\ O_4$, the net magnetic moment is due only to the $Co^{2+}_{oct}$ because the $Fe^{3+}$ moments cancel out. Bulk CFO is inverse (x=1) but the degree of inversion depends on the synthesis method [6,7].

In our previous work [8], lower saturation magnetization was reported in the as-grown monolithic CFO film and BFO-CFO nanocomposite compared to the bulk value, which is due to a Co-rich composition, *i.e.* $Co_{1.2}Fe_{1.8}O_4$ instead of $CoFe_2O_4$. Given the fact that the same CFO target was used for the growth of YFO-CFO composition in this work, a Co-rich composition is also expected. For charge balance the excess Co would be present as $Co^{3+}$. The $Co^{2+}$ is preferentially present on the octahedral sites due to its larger ionic radius. [9] It is therefore reasonable to assume that any excess Co observed at the tetrahedral sites by the STEM (Supplementary Fig. 9b) are +3 valence states. The formula of $Co_{1.2}Fe_{1.8}O_4$ can then be written $[Co^{3+}_{\delta}Fe^{3+}_{1-\delta}]_{tet}\ [Co^{2+}_{1}\ Co^{3+}_{0.2-\delta}\ Fe^{3+}_{0.8+\delta}]_{oct}\ O_4$ where $\delta$ (=0 to 0.2) represents the distribution of the excess Co between the tetrahedral and octahedral sites. Considering the (zero T spin-only) magnetic moments of $Fe^{3+}$ ($d^5$, $5\mu_B$), $Co^{3+}$ ($d^6$, $4\mu_B$) and $Co^{2+}$ ($d^7$, $3\mu_B$), the net moment (2.8-3.2$\mu_B$) is lower that that of the bulk fully inverse spinel $[Fe^{3+}]_{tet} [Co^{2+} Fe^{3+}]_{oct} O_4$ ($4\mu_B$).

When an external magnetic field is applied parallel to the YFO-CFO nanocomposite film ($x$ direction), the CFO pillars undergo a certain deformation due to the negative magnetostrictive coefficient, which is then transferred to YFO through the coherent interfaces and thus generates an internal electric field ($E_i$) which can affect the ferroelectric behavior of YFO. This is the general mechanism of strain-mediated ME coupling in nanocomposite. The $E_i$ caused by external magnetic felds via the ME effect is the key reason for the asymmetric local polarization switching curves [10]. Due to the presence of $E_i$, a larger negative external electric field is required to reverse the positive local polarization, leading to a larger negative coercive field. On the contrary, a smaller positive external field can make the negative local polarization reverse, *i.e.*, the positive coercive field is smaller (Supplementary Fig. 10c). Also due to the presence of $E_i$, the amplitude of piezoresponse is also enhanced upon a positive voltage than negative (Supplementary Fig. 10d).

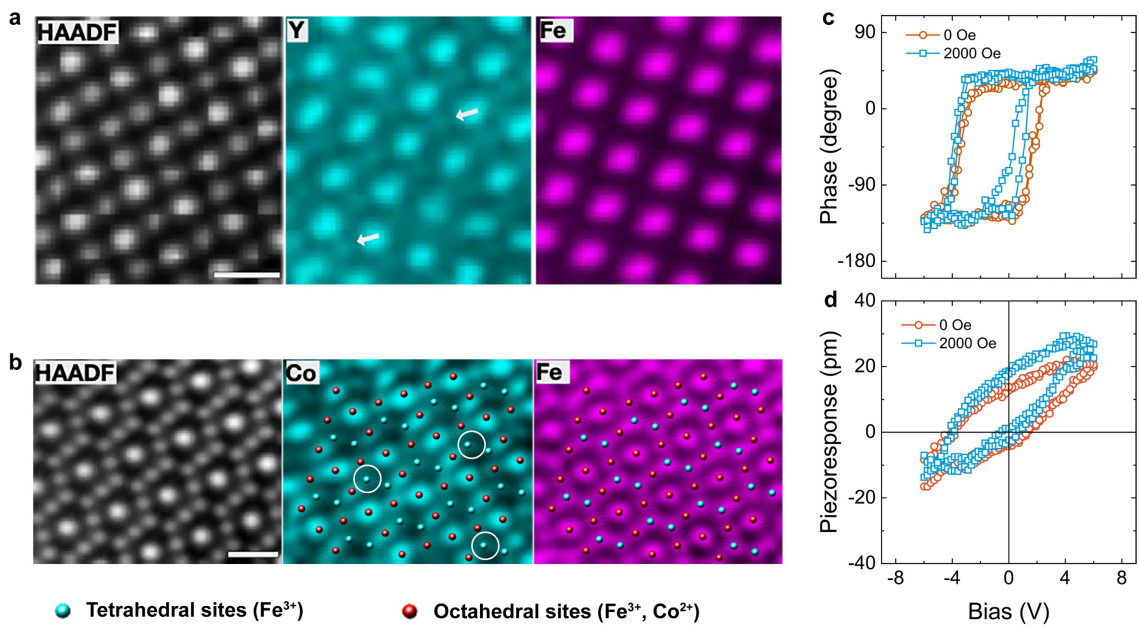

● Tetrahedral sites ($Fe^{3+}$)   ● Octahedral sites ($Fe^{3+}$, $Co^{2+}$)

**Supplementary Fig. 10 │ Atomic structure and ferroelectric properties of YFO-CFO nanocomposite film**. **a**, **b**, HADDF STEM images and STEM EDS elemental mapping collected on YFO (**a**) and CFO (**b**) components, respectively. Scale is 500 pm. A schematic of the CFO lattice is superposed on the EDS map of CFO. White circles indicate existence of Co on tetrahedral Fe sites which is prohibited in perfect spinel CFO crystal. **c**, **d**, *in situ* SS-PFM characterizations include the phase loops (**c**) and the piezoresponce hysteresis (**d**) with and without external magnetic field applied.

# References

1. Hayes, J. R. & Grosvenor, A. P. An X-ray absorption spectroscopic study of the electronic structure and bonding of rare-earth orthoferrites. *J. Phys. Condens. Matter.* **23**, 465502 (2011).

2. Zhou, Z. et al. Ferroelectric domains and phase transition of sol-gel processed epitaxial Sm-doped $BiFeO_3$ (001) thin films. *J. Materiomics* **4**, 27-34 (2018).

3. Zayak, A. T., Huang, X., Neaton, J. B. & Rabe, K. M. Structural, electronic, and magnetic properties of $SrRuO_3$ under epitaxial strain. *Phys. Rev. B* **74** 094104 (2006).

4. Seo, H., Posadas, A. & Demkov, A. A. Strain-driven spin-state transition and superexchange interaction in $LaCoO_3$:Ab initio study. *Phys. Rev. B* **86** 014430 (2012).

5. Zhang, Y., Sun, J., Perdew, J. P. & Wu, X. Comparative first-principles studies of prototypical ferroelectric materials by LDA, GGA, and SCAN meta-GGA. *Phys. Rev. B* **96**, 035143 (2017).

6. A. S. Vaingankar, B. V. Khasbardar, and R. N. Patil, X-ray spectroscopic study of cobalt ferrite, *J. Phys. F: Met. Phys.* **10**, 1615–1619 (1980).

7. G. A. Sawatzky, F. Van Der Woude, and A. H. Morrish, Mössbauer Study of Several Ferrimagnetic Spinels, *Phys. Rev.* **187,** 747 (1969).

8. Zhang, C. et al. Thermal conductivity in self-assembled $CoFe_2O_4$/$BiFeO_3$ vertical nanocomposite films. *Appl. Phys. Lett.* **113**, 223105 (2018).

9. Hou, Y. H. et al. Structural, electronic and magnetic properties of partially inverse spinel $CoFe_2O_4$: a first-principles study. *J. Phys. D: Appl. Phys.* **43** 445003 (2010).

10. Miao, H., Zhou, X., Dong, S., Luo, H. & Li, F. Magnetic-field-induced ferroelectric polarization reversal in magnetoelectric composites revealed by piezoresponse force microscopy. *Nanoscale* **6**, 8515-8520 (2014).